\documentclass[twocolumn,showpacs,preprintnumbers,amsmath,amssymb,prb]{revtex4}

\usepackage{graphicx}
\usepackage{dcolumn}
\usepackage{bm}
\usepackage{bbm}
\usepackage{psfrag}
\usepackage{multirow,amssymb,amsbsy,amsmath}
\usepackage{stmaryrd}

\newcommand{\I}{\textup{i}}
\newcommand{\E}{\textup{e}}
\newcommand{\D}{\textup{d}}
\newcommand*{\Haver}[1]{\mathopen{\llbracket} #1 \mathclose{\rrbracket}}
\newcommand{\vek}[1]{\bm{#1}}
\newcommand{\dd}{\text{d}}
\newcommand{\dod}[2]{\frac{\dd #1}{\dd #2}}

\newcommand{\Funktion}[2]{#1\kern-0.2em\left(#2\right)}

\newcommand{\expfkt}[1]{\Funktion{\exp}{#1}}

\newcommand{\trtxt}[2][]{\text{Tr}_{#1}\{#2\}}
\newcommand*{\bra}[1]{\mathopen{\langle}#1\mathclose{|}}
\newcommand*{\ket}[1]{\mathopen{|}#1\mathclose{\rangle}}

\newcommand{\refsec}[1]{Sect.~\ref{#1}}
\newcommand{\reffig}[1]{Fig.~\ref{#1}}

\newcommand{\comutxt}[2]{[#1,#2]}
\newcommand{\kBolz}{k_{\text{B}}}

\begin{document}

\preprint{APS/123-QED}

\title{Limited validity of the Kubo formula\\
 for thermal conduction
in modular quantum systems}

\author{Jochen Gemmer}%
\affiliation{Physics Department, University of Osnabr\"uck, %
             Barbarastr.\ 7, 49069 Osnabr\"uck, Germany}%
\email{jgemmer@uos.de}%
\author{Mathias Michel}
\affiliation{Institute of Theoretical Physics I, University of Stuttgart, %
             Pfaffenwaldring 57, 70550 Stuttgart, Germany}%
\author{Robin Steinigeweg}
\affiliation{Physics Department, University of Osnabr\"uck, %
             Barbarastr.\ 7, 49069 Osnabr\"uck, Germany}%

\date{\today}

\begin{abstract}
  The Kubo formula describes a current as a response to an external field. 
  In the case of heat conduction there is no such external field. We analyze 
  why and to what extend it is nevertheless justified to describe heat 
  conduction in modular quantum systems by the Kubo formula. ``Modular'' 
  we call systems that may be described as consisting of weakly coupled 
  identical subsystems. We explain in what sense this description applies 
  to a large class of systems. Furthermore, we numerically evaluate the  
  Kubo formula for some finite modular systems. We compare the results with 
  data obtained from the direct numerical solution of the corresponding time-dependent Schr\"odinger equation.
\end{abstract}


\maketitle

%
%

%
%

\section{Introduction}
\label{intro}

There are essentially two reasons that can cause a current of particles like,
e.g., electrons through solids: Either there is an electric field dragging the 
electrons in some direction or the electron density is spatially non-uniform, 
i.e., there is a density gradient causing the electrons to diffuse. For systems
featuring normal transport the currents in those two cases are supposed to be 
determined by 
\begin{subequations}
  \label{eq:1}
  \begin{equation}
    \label{eq:1a}
    \vek{j}=  L_{F}\vek{F}\,,
  \end{equation}
  \begin{equation}
    \label{eq:1b}
    \vek{j}= -L_D \vek{\nabla} \rho\,,
  \end{equation}
\end{subequations}
where $\vek{j}$ is the current, $\vek{F}$ the external force, $\rho$ the spatial 
density and the $L$'s are the pertinent response coefficients. As will be briefly 
outlined below the derivation of (\ref{eq:1a}) from the corresponding scenario is 
rather straightforward and leads to the Kubo formula\cite{Kubo1957,Kubo1957II} (KF). A direct derivation of 
(\ref{eq:1b}) from its underlying scenario seems to be significantly more subtle. 
This is a crucial point for the analysis of heat conduction since, although
being rather similar to the above case otherwise, the scenario corresponding to 
(\ref{eq:1a}) does not exist for heat transport. There simply is no external field 
which could exert a force on heat, heat is always driven by an energy density 
(temperature) gradient as described by (\ref{eq:1b}). Nevertheless, as will also be 
sketched below, there have been attempts to apply the derivation of (\ref{eq:1a}) 
to thermal conduction which eventually implies the application of the KF also in this 
case \cite{Luttinger1964,Kubo1957,Kubo1957II,Kubo1991}.

Let us briefly recall the derivation of the KF. An external field gives rise to an 
additional energy $H'=\int\rho U \D V$ with $-\nabla U=\vek{F}$. This term is 
routinely treated as a perturbing addend to the Hamiltonian of the system yielding an
expression for the induced current. In this expression $\hat{H}'$ itself no longer 
appears only its derivative with respect to time which in the case of a (spatially 
constant) external field reads $\dot{H}'=- \vek{j}\vek{F}$. Thus one obtains a
relation of the form (\ref{eq:1a})\cite{Kubo1957}. $ L_{\vek{F}}$ is given by the Kubo 
formula and reads 
\begin{equation}
  \label{eq:2}
  L(\omega ) = \frac{1}{V}\int_0^{\infty}\D t\, \E^{-\I \omega t} \int_0^{\beta}\D\tau\, 
  \trtxt{\hat{\rho}_0\,\hat{j}(0)\,\hat{j}(t+\I\tau)},
\end{equation}
where $\hat{\rho}_0$ is the Gibbsian equilibrium state, $V$ the volume and $L(\omega)$ is the response 
coefficient which describes the conductivity at frequency $\omega$.

In the case of heat conduction no additional energy arises from the internal force 
(temperature gradient) and thus the stimulus of the current cannot be incorporated 
into the Hamiltonian. Hence the above line of reason does not apply. To nevertheless 
justify a Kubo-type formula for thermal conductivity basically two types of arguments 
are brought forth: \\
i) {\em The hypothesis of local equilibrium} \\
The state that one gets by boldly writing down a Gibbs state with a spatially 
non-uniform temperature that varies little ($\Delta T(\vek{x})$) around some mean 
$\beta_0=1/T_0$ reads
\begin{equation}
 \label{eq:3}
 \hat{\rho}_{\text{leq}}
 =Z^{-1}\text{exp}\Big(-\beta_0
        \int\D\vek{x}\,\frac{1-\Delta T(\vek{x})}{T_0}\,\hat{h}(\vek{x})\Big)
\end{equation}
($Z$ being the partition function, $\hat{h}(\vek{x})$ the local energy density and $T_0$ respectively $\beta_0$ some mean temperature) and is 
called a local equilibrium state (cf.~Ref.\cite{Lepri2003,Zubarev1974}). Its physical significance is somewhat vague since it 
is not a real equilibrium state. However, adding a ``pseudo-perturbation'' of the form
$\hat{H}_{\text{ps}}'=-\int\D\vek{x} \Delta T(\vek{x}) \hat{h}(\vek{x})/T_0$ to the system's 
Hamiltonian and calculating the Gibbs state at uniform $\beta_0$ of the perturbed
system formally yields the above local equilibrium state (\ref{eq:3}). Thus this term 
is said to somehow model the effect of the internal temperature gradient, although the 
local equilibrium state features no current. Nevertheless, proceeding like described above 
and taking $\hat{H}_{\text{ps}}'$ for an external perturbation yields a transport coefficient as 
given by (\ref{eq:2}) (just multiplied by $\beta_0$) with a thermal current $\hat{j}$ 
defined by $\dot{h}+\nabla j=0$.\\
ii) {\em The entropy production argument} \\
If, e.g., an electrical current in some conducting solid runs along an electric field, 
potential energy arising from this field, $H'$, is converted to heat, $Q$. Thus one has 
$-\dot{H}'=\dot{Q}$. Hence the entropy production is $\dot{S}\geq\dot{Q}/T$. On the other 
hand entropy production is assumed to be of the form $\dot{S}=\vek{J}\cdot\nabla T/T^2$ 
(see Refs.\cite{Glansdorff1971,Groot1962,Mahan1981,Forster1975}). Combining these equations and boldly replacing the ``$\geq$'' by an ``$=$'' 
yields $\dot{H}'=-\vek{J}\cdot\nabla T/T$. This can formally be incorporated in the 
derivation of the KF outlined above and yields the same result as the hypothesis of 
local equilibrium. The crucial shortcoming of this argument is that if heat flows along 
a temperature gradient, obviously no heat is produced, only entropy. Thus, in this case 
one would definitely have to consider a ``$>$'' and then this argument yields no concise 
result.

Thus it has often been pointed out that those concepts do not provide a rigorous justification 
of the KF for thermal conduction \cite{Lepri2003,Garrido2001,Bonetto2000,Saito2002,Michel2003,Michel2004,Michel2005} which remains an open question. Furthermore, the
KF has been counter-checked only for few concrete systems, see Refs.\cite{Allen1993,Feldman1993}. Despite those conceptual problems the application of the KF to heat conduction 
is today a standard technique \cite{Zotos1997,Heidrich2003,Kluemper2002}. Thus the main 
intend of the paper at hand is to give a derivation of a Kubo-type formula for thermal 
conduction which is more convincing, thereby also pointing out the limits of its validity.

%
%

\section{Kubo formula for finite quantum systems}

\label{kubo}

Before we proceed with our main line of thought in \refsec{moq} we shortly comment on another 
difficulty that arises if one wants to compute the transport coefficients of finite quantum 
systems from the KF. This point deserves consideration, since, in general, an infinite system 
cannot be treated numerically, thus in many cases a transport coefficient of an (periodic) 
infinite system can only be computed from analyzing a finite piece of the infinite system (cf.~Ref.\cite{Heidrich2005}).

Of course, a key question is whether or not the transport is normal at all. For infinite 
systems the coefficient is believed to take the form $L(\omega)=D_{dr}\delta(0)+\kappa(\omega)$. 
$D_{dr}$ is called the Drude weight and whenever it is nonzero for infinite systems the transport 
is assumed to be ballistic. If it is zero, the normal (diffusive) conductivity is supposed 
to be given by $\kappa(0)$, $\kappa(\omega)$ being a smooth function without any singularities 
\cite{Heidrich2005, Heidrich2003}. However, this distinction is problematic if the KF
is evaluated on the basis of a finite quantum system. In this case $L(\omega)$ consists of a sum 
of delta peaks at different frequencies without any non-singular contribution \cite{Heidrich2005,Heidrich2003}. So, technically speaking, one cannot find normal transport in a finite quantum 
system, it is either ballistic or none. There are different ideas on how the transition to 
infinite systems could produce normal transport. Sometimes the singularity at $\omega=0$ is 
assumed to broaden (``imaginary broadening'') but maintain its weight such that $L(0)=D_{dr}/\tau$ is 
non-singular, where $\tau$ is some inverse width of the Drude-peak which is hard to evaluate 
from a closed finite system \cite{Heidrich2005}. In other approaches $L(\omega)$ is averaged over small 
frequency intervals $\delta \omega$ and the resulting smooth function is extrapolated down to 
$\omega=0$ in order to determine the normal conductivity \cite{Heidrich2005,Heidrich2005II}. 
The result depends, of course, on how $\delta \omega$ is chosen. In general it appears to be 
difficult to determine transport type and conductivity merely from the Hamiltonian of a finite
piece of an infinite system without any further assumptions. 

Thus,
except for demonstrating the limited validity of the KF, the paper at
hand also suggests a consistent method to infer the heat conduction behavior of
an infinite system from analyzing an adequate finite piece of it.

%
%

\section{Modular quantum systems and diffusive dynamics}

\label{moq}

We consider systems consisting of identical many-level subunits which are weakly coupled by 
identical next-neighbor interactions. We call those systems ``modular''. For simplicity, we 
analyze chains and rings of that kind. Their Hamiltonians may be denoted as
\begin{equation}
 \label{eq:4}
 \hat{H} = \sum_{\mu=1}^{N} \hat{h}(\mu) + \sum_{\mu=1}^{N-1} \hat{V}(\mu,\mu+1) \;,
\end{equation}
where $\hat{h}(\mu)$ is the local Hamiltonian of some subunit $\mu$, $N$ the total number of 
subunits and $\hat{V}(\mu,\mu+1)$ represents a next-neighbor interaction between the subsystems 
$\mu$ and $\mu+1$ (Schematic examples of such modular systems are depicted in \reffig{fig:1} 
and \reffig{fig:8}). The one-dimensional character is not crucial here. Everything derived 
below can be generalized straightforwardly to modular systems featuring arbitrary multi-dimensional
``net-structures''. So what sort of physical systems are modular systems? First of all interacting 
nano-structures like arrays of quantum dots, etc., might fit this
scheme. But modular structures may also be achieved by operationally coarse graining periodic (or also slightly 
disordered) systems like spin chains, crystals, etc. in modules such
that each module containins many
elementary cells. The 
interactions between the modules then represent the couplings. Since interactions are typically 
rather short ranged, increasing the ``grain size'' will eventually result in a description in 
which only adjacent modules are coupled. Furthermore, these next neighbor couplings will become 
weaker, such that they finally might be considered as weak. This ``weak coupling'', as well as 
other criteria which determine whether or not there is regular transport, will be given more 
precisely below. They may all depend on the choice of the grain size. This reflects the fact 
that regular thermodynamic behavior may only be expected for a spatially coarse enough
description. Whether or not the criteria for regular transport are fulfilled at some grain 
size has to be investigated for any system individually. However, the description of a system 
as a modular system should be possible for very many systems. Thus, in order to keep everything 
as general as possible, we do not specify our systems in much more detail than already given by
(\ref{eq:4}) (not even the numerical examples in \refsec{mod}). But the concrete application of 
the results at hand to, e.g., spin chains is under way.

What does diffusive transport mean in the context of modular systems? Diffusive heat transport 
is defined by Fourier's law which is essentially given by (\ref{eq:1b}). For modular systems 
we replace the spatial gradient by the energy difference between two adjacent subunits, i.e., 
the current $j(\mu)$ from subunit $\mu$ to subunit $\mu +1$ should obey $j(\mu) = D(h(\mu)-
h(\mu+1))$. A discrete form of the continuity eq.\ for modular systems reads $\dot{h}(\mu)=
j(\mu-1)-j(\mu)$. Combining those eqs.\ yields, e.g., for a chain
\begin{align}
 \label{eq:4a}
 \dod{h_1}{t} &= D(h(2) - h(1))\;, \nonumber \\
 \dod{h_{\mu}}{t} &= D(h(\mu-1) + h(\mu+1) - 2h(\mu))\;,\\ 
 \dod{h_N}{t} &= D(h(N-1) - h(N))\;.\nonumber 
\end{align}
(This may be viewed as a discrete form of the time-dependent version of Fourier's law: 
$\dot{h}(\vek{x})=(\kappa/c)\Delta h(\vek{x})$.) If the motion of energy through the closed 
system (which is entirely controlled by its Hamiltonian and the Schr\"odinger eq.) can be 
described by the above set of eqs., the thermal transport is diffusive. The conductivity is 
obviously related to $D$.

Our paper is roughly organized as follows: We consider the dynamics of a modular system 
without any external forces, starting from a local equilibrium state as given by (\ref{eq:3}). 
We find that, under various conditions on the model parameters, the motion of the energy can be 
described by (\ref{eq:4a}) for a short first time-step (cf.~\refsec{local}), where $D$ is 
essentially determined by the KF. After that time-step the system is unfortunately no longer 
in a local equilibrium state. It can, however, be shown that almost all states sharing some 
crucial properties with the local equilibrium state will also give rise to energy dynamics 
in accord with (\ref{eq:4a}) for another time-step. That means, fully evolved local equilibrium is 
dispensable. Thus one can, iterating the result for short time-steps, conclude that 
(\ref{eq:4a}) provides a correct description for all times (cf.~\refsec{ham}) and thus justify 
the application of the KF. Eventually we evaluate the KF for some concrete finite models and
compare the results with the energy dynamics obtained from a direct numerical integration of 
the corresponding time-dependent Schr\"odinger eqs. (cf.~\refsec{mod}).

%
%

\section{Definition of local energy currents in modular quantum systems}

\label{def}

Consider a Hamiltonian as given by (\ref{eq:4}). In the paper at hand we define the local 
energy operator at site $\mu$ simply as $\hat{h}(\mu)$ rather than $\hat{h}(\mu) + 
(\hat{V}(\mu,\mu-1)+\hat{V}(\mu,\mu+1))/2$. This way the sum of all local energies does not 
represent the total energy and is not a strictly conserved quantity. But the local energy 
operators are defined on strictly separated subspaces of the product space on which the full 
system is defined. Excluding the interactions from the local energies makes a consistent partition 
of the system into mutually disjoint (smallest) subunits possible, such that on each subunit a 
local (equilibrium) state may be defined independently. This would be impossible if one included 
the interactions in the local energies. However, if one wants to define an energy current based 
on the evolutions of those local energies, the sum of the local energies has to be at least
approximately conserved, i.e., the part of the full energy associated with the local energies, 
$\trtxt{\hat{\rho}\,\hat{h}(\mu)}$, has to be much bigger than the part associated with the 
interactions, $\trtxt{\hat{\rho}\,\hat{V}(\mu,\mu+1)}$. This eventually means that the interactions 
have to be \emph{weak}.

For conserved quantities the discrete continuity equation $\dot{h}(\mu)=j(\mu-1)-j(\mu)$ 
suggests the definition of a local current operator \cite{Heidrich2005, Michel2004, Michel2005} via
($\hbar=1$)
\begin{align}
  \label{eq:5}
  \dod{}{t}\hat{h}(\mu)
  &=\I\comutxt{\hat{H}}{\hat{h}(\mu)}\\
  &=\underbrace{\I\comutxt{\hat{V}(\mu-1,\mu)}{\hat{h}(\mu)}}_{\hat{j}(\mu-1)}
   +\underbrace{\I\comutxt{\hat{V}(\mu,\mu+1)}{\hat{h}(\mu)}}_{-\hat{j}(\mu)}
   \;.\notag
\end{align}
However, rewriting the above equation for the temporal change of $\hat{h}(\mu+1)$ produces another 
expression for the local current $\hat{j}(\mu)=\I\comutxt{\hat{V}(\mu,\mu+1)}{\hat{h}(\mu+1)}$. This
can only be consistent if
\begin{equation}
  \label{eq:6}
  \comutxt{\hat{V}(\mu,\mu+1)}{\hat{h}(\mu)+\hat{h}(\mu+1)}\approx 0\;.
\end{equation}
But this expression can be interpreted as the temporal change of the sum of two adjacent local 
energies, if only the interaction between the respective subunits was present. According 
to the weak coupling precondition, this sum of local energies is approximately conserved. Or at 
least its fluctuations are small compared to its mean value as long as the energy contained in 
the subunits is not too small. This means in particular that (\ref{eq:6}) may safely be assumed as 
long as the temperature is not too small (we will come back to that condition later). Thus a
sort of symmetrized local current is defined by
\begin{equation}
  \label{eq:7}
  \hat{j}(\mu) = \I\comutxt{\hat{V}(\mu,\mu+1)}{\hat{S}}\;, 
  \quad 
  \hat{S}= \frac{1}{2}\big(\hat{h}(\mu+1)-\hat{h}(\mu)\big)\;.
\end{equation}
The local current operator is strictly defined within the product space spanned by the 
corresponding adjacent subunits. Its expectation value is determined only by the reduced state 
of those adjacent subunits. Thus, within this framework, the relation between current and 
temperature difference may be determined on the basis of a reduced system consisting of only two 
interacting subunits. In this reduced system the current is simply the temporal change of
$\hat{S}$. And $\hat{S}$ may be interpreted as the operator for the skewness of the energy 
distribution between the two subunits.

%
%

\section{Local energy currents as a response to local temperature gradients}
\label{local}

We now investigate the relation between temperature difference $\Delta
T$ and the short time behavior of the energy current $j$. Inspite of having commented 
in \refsec{intro} on the local equilibrium state in a rather critical way we now analyze the short 
time dynamics, i.e., the formation of a current of an initial local equilibrium state as 
given by (\ref{eq:3}). Since we are not deducing any pseudo potentials here, this does not imply 
the application of the hypothesis of local equilibrium in the sense described in \refsec{intro}. 
However, if one starts with a local equilibrium state and considers only a short time step,
the question arises whether the system will be in (another) local equilibrium state after this 
time step? The answer is no, but this issue is addressed in detail in \refsec{ham}. For the moment we 
thus simply consider the local equilibrium state $\hat{\rho}(T,\Delta T)$ which is defined 
for the two coupled subunits which form our reduced system (for simplicity, those are only 
labeled ``$1$'' and ``$2$'' in the following):
\begin{equation}
  \label{eq:8}
  \hat{\rho}(T,\Delta T)\approx
  \hat{\rho}_0(\hat{1}+\frac{\Delta T}{T^2}\hat{S})\;, 
  \quad
  \hat{\rho}_0:=\frac{\expfkt{\frac{-(\hat{h}(1)+\hat{h}(2))}{T}}}{Z^2(T)}\;.
\end{equation}
($Z$ is the partition function of one subsystem and we chose units of temperature and energy as
$\kBolz=1$, $\hbar=1$.) Here $\hat{\rho}_0$ is obviously a ``global'' equilibrium state with
both subunits at the same temperature $T$.
Unfortunately, the current for this state vanishes, i.e., $\trtxt{\hat{\rho}(T,\Delta T)\,\hat{j}}=0$ 
(if the partial traces of $\hat{V}$ with respect to the subunits vanish, which can be demanded without 
loss of generality).

Thus one has to proceed in a slightly different way: We analyze an approximate short time evolution 
of the expectation value 
\begin{equation}
  \label{eq:9}
  S(\tau) = \trtxt{\hat{\rho}(T,\Delta T)\,\hat{S}(\tau)}
\end{equation}
and consider it's derivative with respect to time for some small but finite  $\tau$. This derivative 
is, according to the Heisenberg equation of motion and the definition in (\ref{eq:7}), the current at 
time $\tau$. The time evolution of the operator $\hat{S}(\tau)$ is computed by means of a truncated 
Dyson series, i.e.,
\begin{align}
  \label{eq:10}
  &\hat{S}(\tau)=\hat{D}^{\dagger}(\tau)\hat{S}\hat{D}(\tau) 
  \quad\text{with}\notag\\
  &\hat{D}(\tau)\approx \hat{1}-\I\hat{U}_1(\tau)-\hat{U}_2(\tau)
\end{align}
and the time evolution operators (see Ref.\cite{Gemmer2004,Gemmer2005I})
\begin{align}
  \label{eq:11}
  \hat{U}_1(\tau) &= \int_0^{\tau}\D\tau'\,\hat{V}(\tau')\;,
  \\
  \label{eq:12}
  \hat{U}_2(\tau) &= \int_0^{\tau}\D\tau'\int_0^{\tau'}\D\tau''
                    \,\hat{V}(\tau') \,\hat{V}(\tau'')\;.
\end{align}
Here the time dependence of the $\hat{V}(\tau)$'s is defined on the basis of an interaction picture, 
i.e., only generated by the local Hamiltonians $\hat{h}(1)+\hat{h}(2)$ rather than the full Hamiltonian. 
This yields in second order approximation
\begin{equation}
  \label{eq:13}
  \hat{S}(\tau)
  \approx \hat{S}+\I\comutxt{\hat{U}_1}{\hat{S}}
  +\hat{U}_1\hat{S}\hat{U}_1-(\hat{S}\hat{U}_2+\hat{U}_2^{\dagger}\hat{S})\;.
\end{equation}
(For simplicity of notation, we suppress the time argument of the $\hat{U}$'s here and in the following.) 
Computing $S(\tau)$ defined in (\ref{eq:9}) by plugging in the operator (\ref{eq:13}) and the state 
(\ref{eq:8}) yields
\begin{equation}
 \label{eq:14}
 S(\tau)=\frac{\Delta T}{T^2}\Big(\trtxt{\hat{\rho}_0\hat{S}^2}
        +\frac{1}{2}\trtxt{\hat{\rho}_0\comutxt{\hat{U}_1}{\hat{S}}^2}\Big)\;,
\end{equation}
where we have exploited $\hat{U}_2+\hat{U}_2^{\dagger}=\hat{U}_1^2$ (which follows from the definitions 
(\ref{eq:11}) and (\ref{eq:12})) and $\comutxt{\hat{\rho}_0}{\hat{S}}=0$ as well as 
$\comutxt{\hat{U}_1}{\hat{\rho}_0}\approx0$. As already mentioned the latter is valid for high enough 
temperatures (remember discussion below (\ref{eq:6})). Realizing, by using the definition of $\hat{U}_1$ 
and (\ref{eq:7}), that
\begin{equation}
  \label{eq:15}
  \comutxt{\hat{U}_1}{\hat{S}}=-\I\int_0^{\tau}\hat{j}(\tau')\D\tau'\;,
\end{equation}
where again $\hat{j}(\tau')$ is defined according to the interaction picture, one can write the derivative 
with respect to time of (\ref{eq:14}) as
\begin{equation}
 \label{eq:16}
 \dod{}{\tau}S(\tau)
 =-\frac{\Delta T}{T^2}\int_0^{\tau}
   \trtxt{\hat{\rho}_0\,\hat{j}(\tau')\,\hat{j}(\tau)}\D\tau'\;.
\end{equation}
Eventually, substituting $t'=\tau-\tau'$ and $t=\tau$
\begin{equation}
 \label{eq:17}
 \dod{}{t}S(t)=j(t)
 =-\frac{\Delta T}{T^2}\int_0^{t}
   \underbrace{\trtxt{\hat{\rho}_0\,\hat{j}(0)\,\hat{j}(t')}}_{=:C(t')}\D t' \;,
\end{equation}
the current is essentially given by an integration over the current auto-correlation function, $C(t')$, 
just like in the KF. Let the timescale on which this correlation typically decays be $\tau_c$. Then, 
if the approximation for $S(t)$ (second order Dyson series) holds for times larger than $\tau_c$ 
(which will be analyzed below), the current will indeed assume a steady value after $\tau_c$ which is 
proportional to $\Delta T$. Thus, first of all, in the case of free heat transport without any external
baths Fourier's law is confirmed for the short-time evolution of a local equilibrium state. The conductivity 
is now defined by
\begin{equation}
  \label{eq:18}
  \kappa=\frac{1}{T^2}\int_0^{\tau_0}C(t)\D t 
  \quad \text{with} \quad 
  \tau_c < \tau_0 <\tau_d\;,
\end{equation}
where $\tau_d$ is the timescale on which the second order approximation for the Dyson series brakes 
down ($\tau_d$ will be evaluated more concretely below). Within this interval the integral should 
not depend much on its upper limit $\tau_0$. Nevertheless, for finite
quantum systems (for which $C(\omega)$ is just a set of peaks)  this value
may differ considerably from 
the one which is produced by letting the upper limit go to infinity. To see this and the relation
to the KF it is instructive to consider $C(\omega)$ and $\bar{C}(\omega)$ defined by 
\begin{align}
  \label{eq:19}
  C(\omega)&=\int_{-\infty}^{\infty}\frac{\E^{-\I\omega t}}{2\pi}\,C(t)\D t\;,
  \\
  \label{eq:20}
  \bar{C}(\omega)&=\frac{1}{\delta \omega}
  \int_{\omega-\frac{\delta \omega}{2}}^{\omega+\frac{\delta \omega }{2}}
  C(\omega)\D\omega\;.
\end{align}
Obviously, $C(\omega)$ is the Fourier transform of $C(t)$ and $\bar{C}(\omega)$ a slightly 
``smeared out'' version of it. Let $\bar{C}(t)$ be the back transform of $\bar{C}(\omega)$. Now, 
as long as $t \ll 2\pi/\delta \omega$, one has $\bar{C}(t) \approx C(t)$. Thus
\begin{equation}  
\label{eq:21}
  \int_{0}^{\tau_0}C(t)\D t\approx
  \int_{0}^{\tau_0}\bar{C}(t)\D t
\end{equation}
as long as $\tau_0 \ll 2\pi/\delta \omega$. Since $\bar{C}(t)$ (other than $C(t)$) does not 
feature any recurrence time if $\bar{C}(\omega)$ is a smooth function
of frequency, we may now 
drive the upper boundary of the second integral to infinity without much changing its value. 
This yields
\begin{equation}
  \label{eq:22}
   \int_{0}^{\tau_0}\bar{C}(t)\D t
  \approx \int_{0}^{\infty}\bar{C}(t)\D t
  =\pi\bar{C}(\omega=0)\;.
\end{equation}
(In the following we denote $\bar{C}(\omega=0)$ simply as $\bar{C}(0)$.) This, however, can 
only hold if $\tau_c \ll 2\pi/\delta \omega$. A rough estimation for $\tau_c$ is given by 
$\tau_c \approx 2\pi/\Delta \omega$ where $\Delta \omega$ is the width of the spectrum of 
$C(\omega)$. Consequently, $\Delta\omega \gg \delta\omega$ has to hold. Of course, $\delta
\omega$ can always be chosen to fulfill this, but if it is chosen too small, $\bar{C}(\omega)$ 
is not necessary a smooth function of frequency. Thus (\ref{eq:22}) eventually holds if 
$\bar{C}(\omega)$ has a reasonably well defined ``peak density'' on a frequency scale small 
compared to its width $\Delta \omega$. In this case $\bar{C}(0)$ does not depend much on 
$\delta \omega$ and there is no need to define $\delta \omega$ with extreme precision.

The KF (also evaluated for a reduced two-subunit system with $V$, as routinely 
done, replaced by the number of contacts between identical subunits, i.e., $V=1$) and its limit 
for $\omega \ll T$, $L'(\omega)$, read in terms of
the correlation function 
\begin{equation}
  \label{eq:23}
  L(\omega)=\pi\frac{1-\E^{-\omega/T}}{T\omega}C(\omega)\;,
  \quad 
  L'(\omega)=\frac{\pi}{T^2}C(\omega)\;.
\end{equation}
This is to be compared with the conductivity as obtained from
(\ref{eq:18}), (\ref{eq:21}), (\ref{eq:22}) which reads
\begin{equation}
  \label{eq:24}
  \kappa=\frac{\pi}{T^2}\bar{C}(0)\;.
\end{equation} 
Obviously, $\kappa$ formally equals $\bar{L}(0)$, the peak density of $L(\omega)$ at frequency 
zero. As mentioned in \refsec{kubo} this quantity has been suggested to compute the conductivity 
of infinite systems from finite models. In this sense and for
$t<\tau_d$ the KF is valid for the computation of 
thermal conductivity, although there is no external potential in this scenario. Note that the 
correlation function in the case of the KF is given by the full Heisenberg dynamics of the system 
based on the full Hamiltonian, whereas $\kappa$ is determined by the correlation function based on 
the interaction picture dynamics, i.e., without taking the interaction into account. However, since 
the whole theory is formulated for the weak coupling limit, this is not going to make a big 
difference as will be numerically demonstrated below (cf. \refsec{mod}).

So far we have always assumed that $\tau_c<\tau_d$. But is that justified? The first order term 
of the Dyson series is $\I\hat{U}_1$. Thus an estimate for the magnitude of the first order term 
which puts weight to the energy subspaces being proportional to their occupation probability is
given by 
\begin{equation}
  \label{eq:25}
  F(t):=\trtxt{\hat{\rho}_0\hat{U}_1^2}\;.
\end{equation} 
Exploiting (\ref{eq:11}) one finds
\begin{equation}
  \label{eq:26}
  \dod{}{t}F(t)\approx \int_0^{t}
  \underbrace{\trtxt{\hat{\rho}_0\,\hat{V}(0)\,\hat{V}(t')}}_{=:C_V(t')}\D t' \;. 
\end{equation} 
Since the decay of the interaction auto-correlation $C_V(t')$ will roughly proceed on the same 
timescale as the current auto-correlation, the temporal change of $F(t)$ will assume a steady value 
after $\tau_c$. Thus, after $\tau_c$, one has $F(t)\approx \bar{C}_V (0)t$, where $\bar{C}_V (0)$ 
is defined analogous to $\bar{C}(0)$ but based on the interaction auto-correlation.  The time for 
a valid approximation according to the Dyson series may now be defined as
\begin{equation}
  \label{eq:27}
  \tau_d = \frac{1}{\bar{C}_{V}(0)}\;,
\end{equation}
a time for which $F(\tau_d)=1$ and consequently the second order approximation definitely brakes down.

If, based on this definition, $\tau_c<\tau_d$ is not fulfilled the current cannot be shown to assume 
a steady value proportional to $\Delta T$ and thus Fourier's law will, in general, not be fulfilled. 
That means the model does not show normal transport. According to (\ref{eq:26}) this happens if the 
coupling  becomes to strong. This fact might be missed by simply evaluating the KF: Since 
$C(\omega)$ completely scales with the square of the overall interaction strength (at least for
$C(\omega)$ evaluated in the interaction picture), the distinction between normal and non-normal 
transport cannot be made by simply looking at the features of $C(\omega)$ (this will be demonstrated 
in more detail in \refsec{mod}). The transport behavior can also become non-normal if the coupling 
becomes too weak. Namely, for $\tau_d \gg 2\pi /\delta \omega$ the current is well predicted by the 
truncated Dyson dynamics for times for which $\bar{C}(t)\approx C(t)$ does not hold any longer. This 
also means that the current cannot be predicted in the described way and thus Fourier's law will 
typically not be fulfilled (this will also be demonstrated in more detail in \refsec{mod}).

%
%

\section{Hilbert space average method and iteration scheme}
\label{ham}

So far we have shown that, under given conditions on the model, for a short time period an energy 
current will flow between the subsystems which is proportional to the local temperature difference 
$\Delta T$. But this has been entirely derived under the assumption that the initial state was a local
equilibrium state of the type $\hat{\rho}(T,\Delta T)$. So what happens after $\tau_d$? It is 
tempting to iterate the procedure, assuming that the state after $\tau_d$ would be again a local
equilibrium state only with some reduced $\Delta T$. Unfortunately, this is rigorously not the case. 
At $t=0$ the initial state $\hat{\rho}(T,\Delta T)$ factorizes. This will for $t>\tau_d$ no longer be the case. The state at 
$t>\tau_d$ features an instantaneous current, $\hat{\rho}(T,\Delta T)$ does not. The full system's 
von-Neumann entropy always equals its initial entropy, but a local equilibrium state with reduced 
$\Delta T$ would feature a higher entropy.

Thus, in the following we abandon the local equilibrium
state. Instead we
show, that essentially almost all states that feature a certain $S$ induce, for a short time-step, a current that is
proportional to $S$. This, of course, results in a state featuring an
accordingly reduced $S$. Assuming that this state also belongs to the
above class, the above short time-step dynamics for $S$ may be
iterated. This yields continuous dynamics for $S$ and hence for the
current. The assumption is only reliable if the above class contains
basically all states in quest.  

The formalism to implement this scheme is called Hilbert space average method (HAM). The key idea of this method is to replace 
the expectation values for observables $\hat{A}$ of actual pure states $\bra{\psi}\hat{A}\ket{\psi}$ (in our case the expectation
value of the energy skewness $\bra{\psi}\hat{S}\ket{\psi}$) by their Hilbert space averages
\begin{equation}
  \label{eq:28}
  \Haver{\bra{\psi}\hat{A}\ket{\psi}}_{\{\bra{\psi}\hat{B}\ket{\psi}=b\}}\;.
\end{equation}
This expression stands for the average of $\bra{\psi}\hat{A}\ket{\psi}$ over all $\ket{\psi}$ that feature
$\bra{\psi}\hat{B}\ket{\psi}=b$ but are uniformly distributed otherwise. Uniformly distributed means invariant 
with respect to all unitary transformations that leave $\bra{\psi}\hat{B}\ket{\psi}=b$ unchanged. The 
replacement of actual expectation values by their Hilbert space
averages is only a justified guess if almost all 
individual $\ket{\psi}$ yield expectation values close to the
Hilbert space average of the observable. It can be shown that this is
the case if the spectral width of $\hat{A}$ is not too large and  $\hat{A}$ is high-dimensional. Full explanation 
of HAM is beyond the scope of this text and can be found in Ref.\cite{Gemmer2003, Gemmer2004}. Here HAM is only 
to be applied.

For the moment let the system be in a pure state. Then the current is given by the temporal change of
$\bra{\psi}\hat{S}(\tau)\ket{\psi}$. Assume that the following set of expectation values is known: 
$\bra{\psi}\hat{P}(\eta)\ket{\psi}=P(\eta)$, where $\hat{P}(\eta)$ is a projector, projecting out the 
subspace which corresponds to an energy interval of width $\Delta \eta$ around $E=\eta$, i.e., 
\begin{equation}
  \label{eq:29}
  \hat{P}(\eta)
  = \sum_{E=\eta-\Delta \eta /2}^{\eta+\Delta \eta /2}\sum_s
    \ket{E,s}\bra{E,s}\;,
\end{equation}
where $E$ are eigenvalues of $\hat{E}:=\hat{h}(1)+\hat{h}(2)$ (sum of local energies) and $s$ eigenvalues 
of $\hat{S}$ (energy skewness). Since, for high enough temperatures, the sum of the local energies is an 
approximately conserved quantity, the $P(\eta)$ are, for large enough $\Delta \eta$, also approximately 
conserved. Assume furthermore that the initial energy skewnesses within all subspaces $\eta$, i.e., 
$\bra{\psi}\hat{P}(\eta)\hat{S}\ket{\psi}=S(\eta)$ are also known. Without taking any further information on 
$\ket{\psi}$ into account the best guess on the evolution of $\bra{\psi}\hat{S}(\tau)\ket{\psi}$ is given by 
the Hilbert space average
\begin{equation}
  \label{eq:30}
  \Haver{\bra{\psi}\hat{S}(\tau)\ket{\psi}}_{\{\bra{\psi}\hat{P}(\eta)\ket{\psi} = P(\eta),\bra{\psi}\hat{P}(\eta)\hat{S}\ket{\psi}=S(\eta)\}}
  = \trtxt{\hat{S}\hat{\alpha}}
\end{equation}
with
\begin{equation}
  \label{eq:31}
  \hat{\alpha}
  = \Haver{\ket{\psi}\bra{\psi}}_{\{\bra{\psi}\hat{P}(\eta)\ket{\psi} = P(\eta),\bra{\psi}\hat{P}(\eta)\hat{S}\ket{\psi}=S(\eta)\}} \; .
\end{equation}
Now what is the above Hilbert space average $\hat{\alpha}$? Any unitary transformation $\hat{G}$ that leaves 
$\hat{P}(\eta)$ and $\hat{P}(\eta)\hat{S}$ invariant has to leave $\hat{\alpha}$ invariant, i.e.,
\begin{align}
  \label{eq:32}
  &\E^{\I\hat{G}}\hat{\alpha}\E^{-\I\hat{G}} = \hat{\alpha} 
  \quad \text{with} \notag\\ 
  &\comutxt{\hat{G}}{\hat{P}(\eta)\hat{S}}
  =\comutxt{\hat{G}}{\hat{P}(\eta)}=0\;.
\end{align}  
This, however, can only be fulfilled if $\comutxt{\hat{G}}{\hat{\alpha}}=0$ for any $\hat{G}$. Furthermore, since the Hilbert 
space average $\hat{\alpha}$ is to be computed under some restrictions (see (\ref{eq:31})), one has the 
following conditions
\begin{equation}
  \label{eq:33}
  \trtxt{\hat{\alpha}\hat{P}(\eta)}=P(\eta)\;, 
  \quad
  \trtxt{\hat{\alpha}\hat{P}(\eta)\hat{S}}=S(\eta)\;.
\end{equation}
According to the invariance properties of $\hat{\alpha}$ and the properties of the operators $\hat{P}(\eta)$ 
and $\hat{P}(\eta)\hat{S}$ (\ref{eq:32}), one may thus write $\hat{\alpha}$ as
\begin{equation}
  \label{eq:34}
  \hat{\alpha}
  = \sum_{\eta}\left(
    P(\eta)\frac{\hat{P}(\eta)}{\trtxt{\hat{P}(\eta)}}
  + S(\eta)\frac{\hat{P}(\eta)\hat{S}}{\trtxt{\hat{P}(\eta)\hat{S}^2}}
    \right)\;.
\end{equation}
Defining $\hat{\rho}_0(\eta):=P(\eta)\hat{P}(\eta)/\trtxt{\hat{P}(\eta)}$, (\ref{eq:34}) may be rewritten as 
\begin{equation}
  \label{eq:35}
  \hat{\alpha}
  = \sum_{\eta}\hat{\rho}_0(\eta)\Big(\hat{1}
  + \frac{S(\eta)}{\trtxt{\hat{\rho}_0(\eta)\hat{S}^2}}\hat{S}\Big)\;.
\end{equation}
By choosing $P(\eta)$ and $S(\eta)$ to be equal to the expectation values of $\hat{\rho}(T,\Delta T)$ for 
the respective operators, i.e.,
\begin{align}
\label{eq:36}  
 P(\eta) &:=\trtxt{\hat{P}(\eta)\hat{\rho}(T,\Delta T)}= \trtxt{\hat{\rho}_0\hat{P}(\eta)} \\
 S(\eta) &:=\trtxt{\hat{P}(\eta)\hat{S}\hat{\rho}(T,\Delta T)}=\frac{\Delta T}{T^2}\trtxt{\hat{\rho}_0(\eta)\hat{S}^2} \nonumber
\end{align}
one gets $\sum_{\eta}\hat{\rho}_0(\eta) \approx \hat{\rho}_0$ and thus $\hat{\alpha}\approx\hat{\rho}(T,\Delta T)$ 
as can be seen from comparison with (\ref{eq:8}). Hence the Hilbert
space average over all pure states featuring 
the same expectation values for $\hat{P}(\eta)$ and $\hat{P}(\eta)\hat{S}$ as $\hat{\rho}(T,\Delta T)$ is 
$\hat{\rho}(T,\Delta T)$ itself. Therefore 
\begin{equation}
  \label{eq:37}
  \bra{\psi}\hat{S}(\tau)\ket{\psi}
  \approx \trtxt{\hat{\rho}(T,\Delta T)\hat{S}(\tau)} \; ,
\end{equation}
i.e.,  any pure state (regardless of whether it is entangled with respect to the subunits or whether its local entropy 
is maximum) featuring the same expectation values for $\hat{P}(\eta)$ and $\hat{P}(\eta)\hat{S}$ as the local equilibrium 
state $\hat{\rho}(T,\Delta T)$ is most likely to yield the same local current as the local equilibrium state. Since 
incoherent mixtures of pure states featuring the same expectation values for $\hat{P}(\eta)$ and $\hat{P}(\eta)\hat{S}$ 
as the local equilibrium state are even closer to the latter then pure states, they are even more likely to induce the 
same current as the local equilibrium state. This holds for the set of  $P(\eta)$, $S(\eta)$ as given by (\ref{eq:36}). 
But how will this set look like after $\tau_0$, i.e., how do the $P(\eta)$ and $S(\eta)$ evolve during the  time $\tau_0$? 
Since the $P(\eta)$ are approximate constants of motion, they remain invariant. For the $S(\eta)$ we find from applying 
the scheme developed in \refsec{local} to $\trtxt{\hat{\alpha}\hat{P}(\eta)\hat{S}(\tau)}$ [cf. (\ref{eq:17}), (\ref{eq:18}), (\ref{eq:24}), (\ref{eq:36})]
\begin{equation}
  \label{eq:38}
  S(\eta, t+\tau_0)
  \approx S(\eta,t) 
  - \underbrace{\frac{\pi\bar{C}(\eta,0)}{\trtxt{\hat{\rho}_0(\eta)\hat{S}^2}}}_{=:2D(\eta)} S(\eta,t)\tau_0\;,
\end{equation}
where $\bar{C}(\eta,0)$ is analogous to  $\bar{C}(0)$ as defined by (\ref{eq:19}) and (\ref{eq:17}) but $\hat{\rho}_0$ 
replaced by $\hat{\rho}_0(\eta)$. If the criteria from \refsec{local} are fulfilled and $\tau_0$ is comparatively short, 
this may now be iterated yielding
\begin{equation}
  \label{eq:39}
  \dod{}{t}S(\eta)=-2D(\eta)S(\eta)\;.
\end{equation}
Thus, in general, any energy subspace has its own energy diffusion coefficient. Therefore it might, 
be impossible to describe the energy diffusion behavior between the subunits entirely by one overall conductivity 
as implied by the KF, even if all the criteria from (\ref{eq:17}) are fulfilled. However, if the $D(\eta)$ 
corresponding to the energy subspaces that are occupied with significant weight and contribute significantly to the 
transport feature similar values, i.e, $D(\eta)\approx D$, one gets only one single diffusion coefficient that then 
reads
\begin{equation}
  \label{eq:40}
  D \approx \frac{\pi\bar{C}(0)}{\trtxt{\hat{\rho}_0\hat{S}^2}}\,.
\end{equation}
(Note that $\sum_{\eta}\bar{C}(\eta,0)=\bar{C}(0)$). In this case one might sum (\ref{eq:39}) over $\eta$ which yields
\begin{equation}
  \label{eq:41}
  \dod{}{t}S=-2DS\;.
\end{equation}
It is straightforward to show that the heat capacity $c$ for one subsystem is given by $c=2\trtxt{\hat{\rho}_0\hat{S}^2}/T^2$. Exploiting 
this and (\ref{eq:40}) one may rewrite (\ref{eq:41}) as
\begin{equation}
  \label{eq:42} 
  \dod{}{t}S=-\frac{2\kappa}{c}S \quad \mbox{or}\quad j(1)=\frac{\kappa}{c}(h(1)-h(2))
\end{equation}
which, generalized to $N$
subsystems, identifying $D=\kappa/c$ and combined with the
discrete continuity eq., yields (\ref{eq:4a}). Hence this 
closes the loop for a microscopic derivation of (\ref{eq:4a}). This essentially means that if the criteria for 
$\tau_0,\tau_c,\tau_d$ from \refsec{local} are fulfilled and the relevant $D(\eta)$ are similar, energy will diffuse 
from subunit to subunit as predicted by Fourier's law.

%
%

\section{Application to models}
\label{mod}
In this Section the previously derived theoretical results are compared with numerical data in the following 
way: For concretely defined models the energy diffusion coefficient $D=\kappa/c$ which appears in
(\ref{eq:4a}) is computed from the KF. Then (\ref{eq:4a}) is solved for some non-uniform initial energy distribution. 
Furthermore, the time-dependent Schr\"odinger eq. is solved numerically for an initial state corresponding to the
above initial energy distribution. From those data the exact evolution of the local energies 
$h(\mu)=\bra{\psi(t)}\hat{h}(\mu)\ket{\psi(t)}$ is computed and compared with the above solution of (\ref{eq:4a}). 
Only if there is good agreement, the system exhibits normal transport that may be characterized by the conductivity 
obtained from the KF.

\begin{figure}
  \centering 
  \includegraphics[width=7cm]{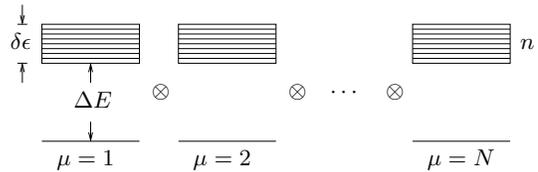}
  \caption{Simple model to analyze diffusive energy transport: $N$
    coupled subunits featuring a single ground state and an excitation
    band of $n$ equally distributed levels.}
\label{fig:1}  
\end{figure}

We introduce two classes of models which are primarily designed to represent modular systems in general rather 
than real physical systems. (For the impact on real physical systems see \refsec{moq}.) The first class of models 
features subunits with non-degenerate ground states, large energy gaps ($\Delta E$) and one, comparatively narrow
energy band ($\delta \epsilon$) each, as depicted in
\reffig{fig:1}. 
Within one band there are $n$ states featuring equidistant level spacing. The next neighbor interactions are defined as
\begin{equation}
  \label{eq:43}
  \hat{V}=\lambda \sum_{i,j,\mu}v(i,j)\hat{P}^+(i,\mu)\hat{P}^-(j,\mu+1)
          +\text{h.c.}
\end{equation}
(h.c. stands for the hermitian conjugate of the previous sum.) Here $\hat{P}^+(i,\mu)$ corresponds to a transition of the $\mu$'th
subunit from its ground state to the $i$'th eigenstate of the band. $\hat{P}^-$ corresponds to the respective downwards transition. 
The $v(i,j)$ are randomly distributed complex numbers normalized to $\sum_{i,j}|v(i,j)|^2/n^2=1$. Thus $\lambda$ sets the overall 
interaction strength. Due to this model design only one energy subspace contributes to transport at all, which is the
``one-excitation subspace'' defined by an overall energy $E$ with $\Delta E \leq E\leq \Delta E+\delta \epsilon$. The dimension 
$M=nN$ of this relevant subspace grows linearly rather than exponentially with the number of subunits. This allows to numerically 
analyze models with high enough $n$ to fulfill all criteria for diffusive transport, but also up to fifteen subunits. Thus those 
models are meant to check whether or not the locally computed conductivity holds for arbitrarily many subunits, as predicted by 
the theory. (For a  ``stand alone'' treatment of such
systems, cf. Ref. \cite{Michel2005II}.)

\begin{figure}
  \centering 
  \includegraphics[width=7cm]{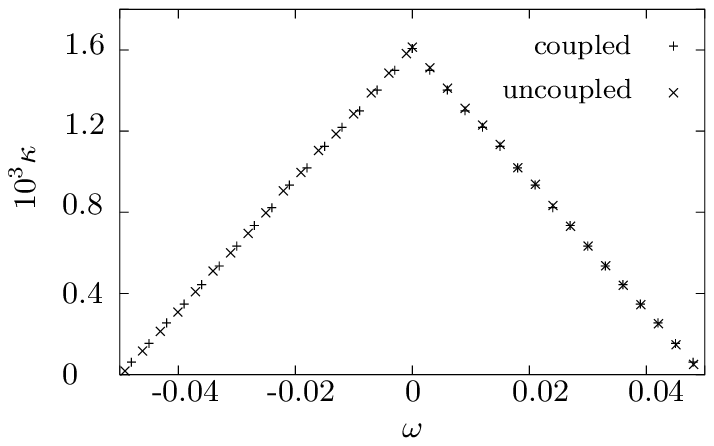}
  \caption{Heat conductivity over frequency for a ringlike system as depicted in \reffig{fig:1} ($N=6$, other parameters see
    text). Crosses refer to the coupled system, as intended by the Kubo formula, and x's to the decoupled system, as suggested 
    in \refsec{local}.}
\label{fig:2}  
\end{figure}

We consider a ring of $N=6$ subsystems with $n=500$, $\lambda =5\cdot 10^{-5}$, $\Delta E =10$ and $\delta\epsilon =0.05$. We 
find that with a frequency averaging interval of $\delta \omega \approx 10^{-3}=\delta\epsilon/50$ all the conditions mentioned 
in \refsec{local} are fulfilled. Since only one energy subspace contributes to transport, none of the difficulties concerning 
different $L(\eta)$ discussed in \refsec{ham} arises. Thus the KF may be evaluated at any temperature to compute $D=\kappa/c$. 
We evaluated $\kappa(\omega)=\bar{C}(\omega)/T^2$ (cf. \refsec{local}) for $T=1.4$ on the basis of the coupled system, 
as intended by the Kubo formula, and on the basis of the uncoupled system, as intended by our argument in \refsec{local}. Both 
results are displayed in \reffig{fig:2}. Obviously, there is a good agreement between the two graphs, it appears to be irrelevant 
whether the correlation function is evaluated on the basis of the coupled or the decoupled system. (It should be mentioned that 
in both cases $C(\omega)$ features finite contributions exactly at $\omega=0$, thus there is a finite Drude peak.) From 
\reffig{fig:2} we find
\begin{figure}
  \centering 
  \includegraphics[width=7cm]{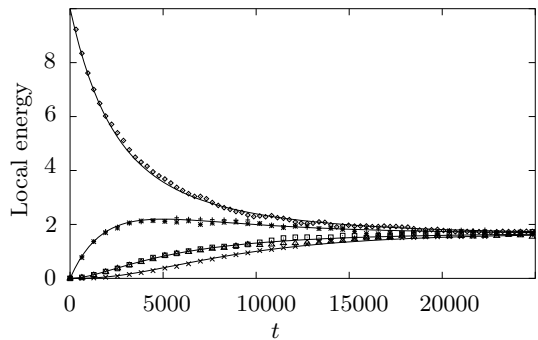}
  \caption{Evolution of the local energies of a weakly coupled system as depicted in \reffig{fig:1}. The initial state
    features one excited subsystem. Displayed are the predictions from the Kubo formula (solid lines) and the complete exact
    solution of the Schr\"odinger equation (points). The figure indicates diffusive transport in accord with the Kubo formula.}
\label{fig:3}  
\end{figure}%
$\kappa=1.6\cdot 10^{-3}$ ($\omega=0$) and calculating the specific heat for one subunit yields $c(T=1.4)=10.5$. This yields 
$D=3.142 \cdot 10^{-4}$. The corresponding solution of (\ref{eq:4a}) for all energy initially concentrated in one subsystem
is shown in \reffig{fig:3}. Furthermore, we solved the Schr\"odinger eq.\ for a corresponding pure initial product state, 
featuring one subunit in a randomly generated state restricted to the excitation band, and all other subunits in their
ground states. The result is also shown in \reffig{fig:3}. Obviously, there is fairly good agreement. We checked rings up to
fifteen subunits and always found good agreement. So far the KF appears to be perfectly valid for heat conduction.

But are the criteria for diffusive transport from \refsec{local} fulfilled? An estimate for the correlation time is given 
by $\tau_c \approx 2\pi/\delta \epsilon \approx 10^2$. From evaluating (\ref{eq:27}) one finds $\tau_d \approx 7\cdot 10^{3}$. 
Thus $\tau_c \ll \tau_d$ obviously holds. Furthermore $2\pi/\delta \omega\approx 6 \cdot10^3$, 
hence $\tau_d \approx 2\pi/\delta\omega$, i.e., the criteria for diffusive transport are fulfilled.

\begin{figure}
  \centering 
  \includegraphics[width=7cm]{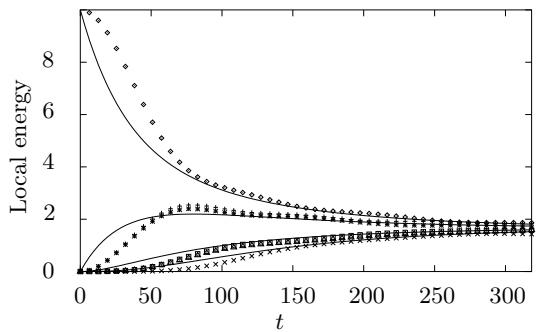}
  \caption{Evolution of the local energies of a strongly coupled system as depicted in \reffig{fig:1}. The deviation of 
  the points (Schr\"odinger equation) from the solid lines (Fourier's law, Kubo formula) indicate the breakdown of 
  diffusive behavior and the validity of the Kubo Formula.}
\label{fig:5}  
\end{figure}

\begin{figure}
  \centering 
  \includegraphics[width=7cm]{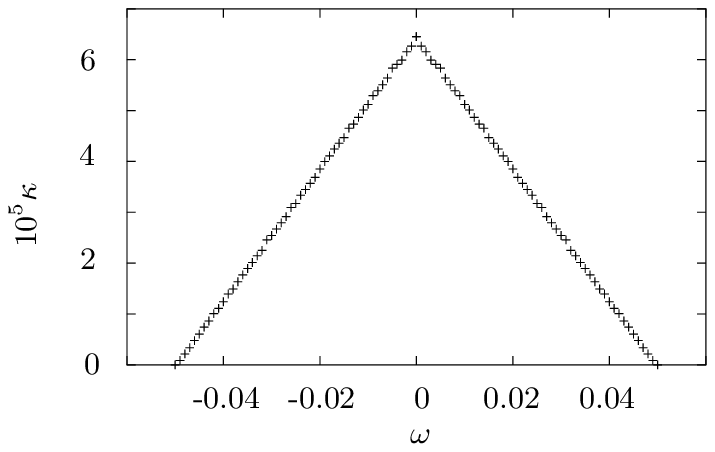}
  \caption{Conductivity over frequency as calculated from the Kubo formula for an extremely weak coupled system as 
  depicted in \reffig{fig:1}.}
\label{fig:6}  
\end{figure}

\begin{figure}
  \centering 
  \includegraphics[width=7cm]{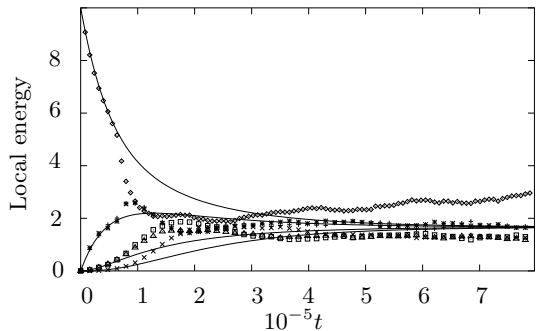}
  \caption{Evolution of the local energies of an extremely weakly coupled system as depicted in \reffig{fig:1}. The 
  deviation of the points (Schr\"odinger equation) from the solid lines (Fourier's law, Kubo formula) indicate the breakdown
  of diffusive behavior and the validity of the Kubo Formula.}
\label{fig:7}  
\end{figure}

If, however, the interaction strength is such that the criteria from \refsec{local} are not fulfilled the transport 
behavior ceases to be diffusive, i.e., the good agreement between the solution of (\ref{eq:4a}) and the solution of the 
Schr\"odinger eq.\ vanishes. For example, for a model like the above one but with $\lambda =  5\cdot 10^{-4}$ one has 
$\tau_d \approx 70$ and $\tau_c \ll \tau_d$ is not fulfilled.
\reffig{fig:5}
\begin{figure}
  \centering 
  \includegraphics[width=7cm]{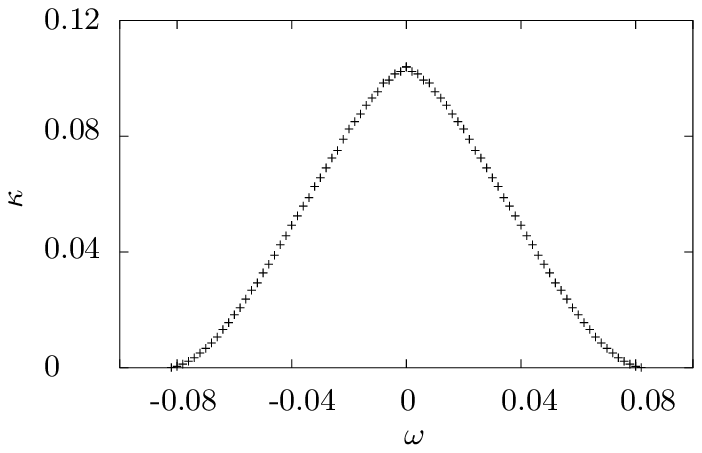}
  \caption{Conductivity over frequency as calculated from the Kubo formula for a strongly coupled system as depicted 
  in \reffig{fig:1}.}
\label{fig:4}  
\end{figure}%
shows the significant deviations of the evolution of the local energies from normal diffusive behavior as described by 
(\ref{eq:4a}). But the graph for $\kappa(\omega)$ as calculated from the Kubo formula (see \reffig{fig:4}) does not 
look essentially different from the above regular case. In particular there is no pronounced singularity at $\omega =0$ 
as expected for the case of ballistic transport. Thus it is not obvious how the general transport behavior is to be 
found from simply evaluating the Kubo formula since there is no $\tau_d$ to be checked. The same is found for extremely 
weak interactions. For example, for a model like the above one but with $\lambda =  10^{-5}$ one has 
$\tau_d \approx 1.8\cdot 10^{5} $ and $ \tau_d \approx < 2\pi/\delta \omega$ is not fulfilled. Consequently, the regular 
transport behavior breaks down (see \reffig{fig:7}). But again, the graph for  $\kappa(\omega)$ as calculated from the 
Kubo formula does not look essentially different from the regular case (see \reffig{fig:6}).

\begin{figure}
  \centering 
  \includegraphics[width=7cm]{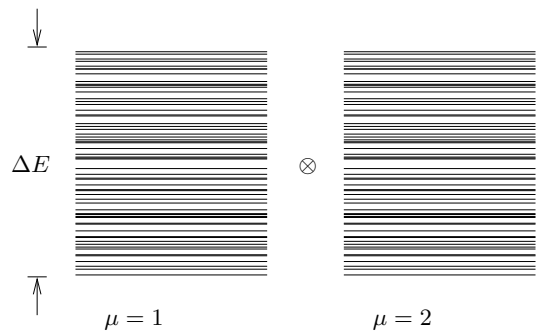}
  \caption{Gapless model to analyze diffusive transport: Two coupled subunits featuring $n$ levels which are uniformly 
  distributed within the energy interval $\Delta E$.}
\label{fig:8}  
\end{figure}

The other model is meant to show that the gapped spectrum of the subunits is dispensable for demonstrating diffusive 
transport. It consists of subunits featuring $n$ eigenstates distributed uniformly within an energy interval
$\Delta E$ as depicted in \reffig{fig:8}.
\begin{figure}
  \centering 
  \includegraphics[width=7cm]{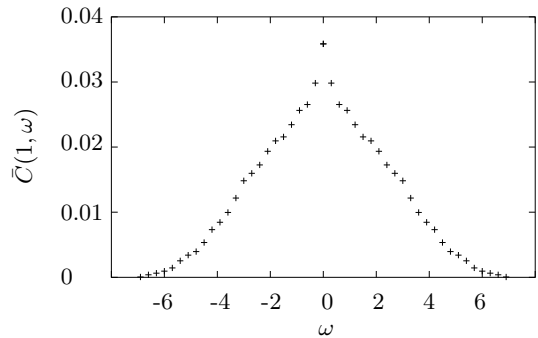}
  \caption{Fourier transform of the current auto-correlation function for the model depicted in \reffig{fig:8} 
  (parameters see text).}
\label{fig:9}
\end{figure}%
\begin{figure}
  \centering 
  \includegraphics[width=7cm]{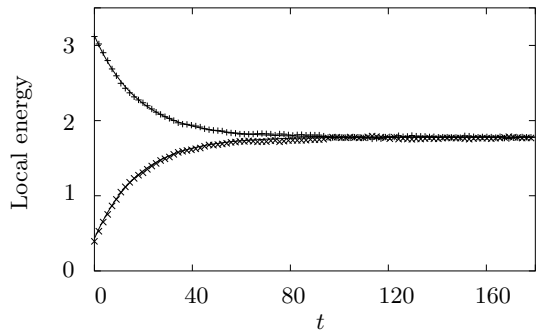}
  \caption{Evolution of the local energies for the model depicted in \reffig{fig:8} for an initial state featuring 
  $T_1\gg \Delta E,\; T_2 \ll \Delta E$. Solid lines refer to the HAM prediction (\ref{eq:4a}), points to the 
  Schr\"odinger equation. The model clearly shows diffusive transport.}
\label{fig:10}  
\end{figure}%
Here the interaction is chosen to be a complex random matrix on the full system's space without any restriction to a
subspace. Nevertheless, the interaction is supposed to be given by $\hat{V}=\lambda \hat{v}$ with $\trtxt{\hat{v}^2}/n^2=1$. 
To keep the problem numerically manageable we consider only two subunits. We specify our model concretely by $N=2$, 
$n=60$, $\lambda =5\cdot 10^{-3}$, $\Delta E =7$. For this case we consider a (``pseudo-thermal'') pure product initial
state. The amplitudes of this state are chosen such that their magnitude squares obey Boltzmann distributions with 
$T_1=40$, $T_2=1$. The phases of the amplitudes are chosen at random. This state has been chosen since for states with 
lower temperature differences it is hard to distinguish relaxation behavior from fluctuations. Since for this model 
various energy subspaces $\eta$ yield different transport types and coefficients $D(\eta)$, the transport behavior for 
this initial state cannot simply be determined by directly applying the Kubo formula. It may, however, be analyzed 
within the framework described in \refsec{ham}. Since $\Delta \eta$ eventually determines the correlation time $\tau_c$
corresponding to the energy subspace $\eta$, one must, as eventually turns out, divide the energy scheme of this model 
in just two subspaces. Thus $\hat{P}(1)$ projects out all states with total energy $E$ lower than $\Delta E$, 
$0 \leq E \leq \Delta E$, and  $\hat{P}(2)$ the states with  $\Delta E < E \leq 2\Delta E$. Based on those definitions 
one finds for the above initial state $P(1)\approx 1$, $P(2)\approx 0$. Thus the relaxation behavior is controlled by 
$D(1)$ as given by (\ref{eq:38}). \reffig{fig:9} shows the corresponding $\bar{C}(1, \omega)$ and we find 
$\bar{C}(1, 0)=0.036$. According to the above definitions we have $\tau_c\approx 2\pi/\Delta E\approx 0.9$. From 
numerically evaluating $\tau_d$ we find $\tau_d \approx 40$ and thus the corresponding condition for normal energy
diffusion $\tau_c \ll\tau_d$ is fulfilled. With $\delta\omega=0.3$ we get $2\pi/\delta\omega \approx 21$ which is of 
the same order of magnitude as $\tau_d$, which also indicates diffusive transport. From numerical evaluation we furthermore 
find $\trtxt{\hat{\rho}_0(1)\hat{S}^2}=2.03$. Those numbers eventually yield $D(1)=0.0139$. Indeed, as displayed
in \reffig{fig:10}, there is good agreement of the solution of (\ref{eq:4a}) based on this $D(1)$ with the full numerical
solution of the corresponding Schr\"odinger equation.

%
%

\section{Summary and conclusion}

\label{sum}

We investigated the energy transport behavior of modular quantum systems, i.e., systems that can be 
described as weakly coupled identical many-level subunits. We argued that this description may apply to a 
large class of systems. For simplicity, we concretely analyzed modular chains and rings with next neighbor 
couplings. Without making any reference to external forces or exploiting the hypothesis of local equilibrium 
we showed that those systems may or may not exhibit normal heat transport, but if they do, the conductivity 
is correctly described by the Kubo formula. This is in accord with Refs.\cite{Allen1993, Feldman1993, Saito2002} 
where the Kubo Formula has been evaluated for concrete systems and the results have been counter-checked by 
either experiments or other theoretical methods. This is furthermore in accord with Refs.\cite{Prosen2000,Li2004,Saito1996,Saito2003,MejiaMonasterio2005} where it is shown that one-dimensional 
systems may or may not exhibit diffusive heat transport.

We also suggested general criteria to decide whether or not such modular systems exhibit normal transport. 
Those criteria are established on the basis of the concrete form of the subunits and their mutual interactions. 
We found, however, that the question cannot be decided only by evaluating the Kubo Formula. To check our 
theoretical results we introduced some examples for concrete, finite, modular, chainlike systems. For those
examples we numerically solved the time-dependent Schr\"odinger equation which yields the energy transport 
dynamics. Those dynamics are in accord with our above mentioned theoretical results.

The results support the view that thermodynamic behavior might, under specific conditions on the system, 
emerge directly from quantum mechanics \cite{GemmerOtte2001,Gemmer2004,Zurek1994}. These conditions do 
not necessarily include a many particle limit.

We sincerely thank F. Heidrich-Meisner for very fruitful discussions. Financial support by the Deutsche 
Forschungsgesellschaft is gratefully acknowledged.


\end{document}